\documentclass[11pt]{article}
\usepackage{amssymb,amsmath,amsfonts}
\usepackage{graphicx}
\usepackage{graphics}
\usepackage{eepic,epsfig}

\begin{document}

\title{Qualitative evolution in $f(R)$ cosmologies}
\author{ R. M. Avagyan, E. V. Chubaryan, G. H. Harutyunyan, A. A. Saharian \\
\textit{Department of Physics, Yerevan State University,}\\
\textit{1 Alex Manoogian Street, 0025 Yerevan, Armenia }}
\maketitle

\begin{abstract}
We investigate the qualitative evolution of $(D+1)$-dimensional cosmological
models in $f(R)$ gravity for the general case of the function $f(R)$. The
analysis is specified for various examples, including the $(D+1)$%
-dimensional generalization of the Starobinsky model, models with polynomial
and exponential functions. The cosmological dynamics are compared in the
Einstein and Jordan representations of the corresponding scalar-tensor
theory. The features of the cosmological evolution are discussed for
Einstein frame potentials taking negative values in certain regions of the
field space.
\end{abstract}

\bigskip

PACS numbers: 03.70.+k, 11.10.Kk, 03.75.Hh

\bigskip

\section{Introduction}

Recent observations of the cosmic microwave background, large scale
structure and type Ia supernovae have provided strong evidence that at
present epoch the expansion of the universe is accelerating \cite{Suzu07}.
Assuming that General Relativity correctly describes the large scale
dynamics of the Universe, this means that the energy density is currently
dominated by a form of energy having negative pressure. This type of a
gravitational source is referred as dark energy. The simplest model for the
latter, consistent with all observations to date, is a cosmological
constant. From the cosmological point of view, a cosmological constant is
equivalent to the vacuum energy in quantum field theory. However, the value
of a cosmological constant inferred from cosmological observations is many
orders of magnitude smaller than the value one might expect based on quantum
field-theoretical considerations. This large discrepancy is one of the
motivations to consider alternative models for dark energy. To account for
the missing energy density, instead of a cosmological constant one could add
a new component of matter, such as quintessence (see \cite{Cope06} for a
review). The latter is modeled by slow-rolling scalar fields. However,
because of very small mass of a scalar field responsible for the
acceleration, it is generally difficult to construct viable potentials on
the base of particle physics.

More recently, it has been shown that suitable modifications of General
Relativity can result in an accelerating expansion of the Universe at
present epoch. These modifications fall into two general groups. The first
one consists of scalar-tensor theories that are most widely considered
extensions of General Relativity \cite{Will93}. In addition to the metric
tensor, these theories contain scalar fields in their gravitational sector
and typically arise in the context of models with extra dimensions
(Kaluza-Klein-type models, braneworld scenario) and within the framework of
the low-energy string effective gravity. In the second group of models, the
Ricci scalar $R$ in the Einstein-Hilbert action is replaced by a general
function $f(R)$ (for recent reviews see \cite{Noji07}-\cite{Clif12}). One of
the first models for inflation with quadratic in the Ricci scalar
Lagrangian, proposed by Starobinsky \cite{Star80}, falls into this class of
theories. An additional motivation for the $f(R)$ theories comes from
quantum field theory in classical curved backgrounds \cite{Birr82} and from
string theories. The recent investigations of cosmological models in the $%
f(R)$ theories of gravity have shown a possibility for a unified description
of the inflation and the late-time acceleration.

$f(R)$ gravities can be recast as scalar tensor theories of a special type
with a potential determined by the form of the function $f(R)$. Various
special forms of this function have been discussed in the literature. In
particular, the functions were considered that realize the cosmological
dynamics with radiation dominated, matter dominated and accelerated epoch.
Unified models of inflation and dark energy have been studied as well \cite%
{Noji11}. In the present paper we consider the qualitative evolution of the
cosmological model for a general $f(R)$ function. The general analysis is
specified for various examples, including the original Starobinsky model. We
have organized the paper as follows. In the next section we present the
action of $f(R)$ gravity in the form of the action of a scalar-tensor theory
in a general conformal representation. Then, the general action is specified
for the Einstein frame with a scalar field having a canonical kinetic term.
The specific form of the scalar field potential is given for various
examples of the $f(R)$ function. The corresponding cosmological model is
described in section \ref{sec:Cosm} and the relations between the functions
in the Einstein and Jordan frames are discussed. The qualitative analysis of
the spatially flat gravi-scalar model is presented in section \ref%
{sec:Qualit}. The phase portraits are plotted for special cases. The main
results of the paper are summarized in section \ref{sec:Conc}.

\section{$f(R)$ gravity as a scalar-tensor theory: Conformal representations
and examples}

\label{sec:fR}

The action in $(D+1)$-dimensional $f(R)$ theory of gravity has the form%
\begin{equation}
S=\int d^{D+1}x\sqrt{\left\vert g\right\vert }\left[ f(R)+L_{m}\left(
g_{ik},\psi \right) \right] ,  \label{SfR}
\end{equation}%
where $L_{m}\left( g_{ik},\psi \right) $ is the Lagrangian density for
non-gravitational matter collectively denoted by $\psi $. It is well known
(see \cite{Noji07}-\cite{Clif12}) that (\ref{SfR}) can be presented in the
form of the action for scalar-tensor gravity. In order to show that we
consider the action%
\begin{equation}
S=\int d^{D+1}x\sqrt{\left\vert g\right\vert }\left[ f^{\prime }(\eta
)(R-\eta )+f(\eta )+L_{m}\left( g_{ik},\psi \right) \right] ,  \label{SfRb}
\end{equation}%
with a scalar field $\eta $. The equation for the latter is reduced to $%
f^{\prime \prime }(\eta )(R-\eta )=0$. Assuming that $f^{\prime \prime
}(\eta )\neq 0$, from the field equation we get $\eta =R$. With this
solution, the action (\ref{SfRb}) is reduced to the original action (\ref%
{SfR}).

Introducing a new scalar field $\varphi $ in accordance with%
\begin{equation}
\varphi =-f^{\prime }(\eta ),  \label{phidef}
\end{equation}%
the action (\ref{SfRb}) is written in the form%
\begin{equation}
S=\int d^{D+1}x\sqrt{\left\vert g\right\vert }\left[ -\varphi R-V(\varphi
)+L_{m}\left( g_{ik},\psi \right) \right] ,  \label{SfRc}
\end{equation}%
with the scalar potential%
\begin{equation}
V(\varphi )=-f(\eta (\varphi ))-\varphi \eta (\varphi ).  \label{PotFR}
\end{equation}%
Here, we have assumed that the function $\varphi (\eta )$, defined by (\ref%
{phidef}), is invertible. The action (\ref{SfRc}) describes a scalar-tensor
theory. In the representation (\ref{SfRc}) the Lagrangian density of the
non-gravitational matter does not depend on the scalar field $\varphi $.
Hence, the representation corresponds to the Jordan frame.

The scalar-tensor theories can be presented in various representations which
are related by conformal transformations of the metric tensor. Let us
consider a general conformal transformation
\begin{equation}
g_{ik}=\Omega ^{2}(\varphi )\tilde{g}_{ik},  \label{Conf}
\end{equation}%
with a sufficiently smooth function\ $\Omega (\varphi )$. Up to total
derivative terms, in the new conformal representation the action takes the
form\
\begin{equation}
S=\int d^{D+1}x\,\sqrt{\left\vert \tilde{g}\right\vert }\left[ -\tilde{F}%
_{R}(\varphi )\tilde{R}+\tilde{F}_{\varphi }(\varphi )\tilde{g}^{ik}\partial
_{i}\varphi \partial _{k}\varphi -\;\tilde{V}(\varphi )+\tilde{L}%
_{m}(\varphi ,\tilde{g}_{ik},\psi )\right] ,  \label{S2}
\end{equation}%
where we have introduced the notations
\begin{eqnarray}
\tilde{F}_{\varphi }(\varphi ) &=&-D\Omega ^{D-1}(\Omega ^{\prime }/\Omega )%
\left[ (D-1)\varphi \Omega ^{\prime }/\Omega +2\right] \,,  \notag \\
\tilde{F}_{R}(\varphi ) &=&\Omega ^{D-1}\varphi ,\;\tilde{V}(\varphi
)=\Omega ^{D+1}V(\varphi ),\;\tilde{L}_{m}(\varphi ,\tilde{g}_{ik},\psi
)=\Omega ^{D+1}L_{m}(\Omega ^{2}\tilde{g}_{ik},\psi ),  \label{FR}
\end{eqnarray}%
and the prime stands for the derivative with respect to $\varphi $.
Qualitative evolution of the models of the type (\ref{S2}) with $\;\tilde{V}%
(\varphi )=0$, arising in higher-loop string cosmology, has been discussed
in \cite{Saha99}.

By choosing the conformal factor as%
\begin{equation}
\Omega (\varphi )=\Omega _{E}\left( \varphi \right) =m_{P}\varphi ^{1/(1-D)},
\label{OmE}
\end{equation}%
where $m_{P}=1/(16\pi G_{D+1})^{1/(D-1)}$ is the Planck mass in $(D+1)$%
-dimensions and $G_{D+1}$ is the corresponding gravitational constant, we
get $\tilde{F}_{R}(\varphi )=m_{P}^{D-1}$. In the corresponding conformal
frame, referred as the Einstein frame, the gravitational part of the action
takes the form of that for $(D+1)$-dimensional General Relativity:
\begin{equation}
S=\int d^{D+1}x\sqrt{\left\vert g_{(E)}\right\vert }\left[
-m_{P}^{D-1}R_{(E)}+F_{\varphi }^{(E)}\left( \varphi \right)
g_{(E)}^{ik}\partial _{i}\varphi \partial _{k}\varphi -V_{(E)}(\varphi
)+L_{m}^{(E)}\left( \varphi ,g_{(E)ik},\psi \right) \right] .  \label{SE}
\end{equation}%
In the Einstein frame, the function in front of the scalar field kinetic
term and the non-gravitational Lagrangian density are related to the
functions in the original action by the formulae%
\begin{equation}
F_{\varphi }^{(E)}\left( \varphi \right) =m_{P}^{D-1}\frac{D\varphi ^{-2}}{%
D-1},\;V_{(E)}(\varphi )=\frac{m_{P}^{D+1}V(\varphi )}{\varphi ^{(D+1)/(D-1)}%
},  \label{FE}
\end{equation}%
and%
\begin{equation}
L_{m}^{(E)}\left( \varphi ,g_{(E)ik},\psi \right) =m_{P}^{D+1}\varphi
^{(D+1)/(1-D)}L_{m}\left( \Omega _{E}^{2}g_{(E)ik},\psi \right) .
\label{LmE}
\end{equation}%
The function $F_{\varphi }^{(E)}\left( \varphi \right) $ has a pole at $%
\varphi =0$. As it has been discussed in \cite{Saha00}, the presence of
singularities in the kinetic function for a scalar field provides an
additional mechanism for the cosmological stabilization of scalar fields.
Note that in a large class of models discussed in \cite{Gala15} the
inflationary predictions for the spectral index and for the tensor-to-scalar
ratio are determined by the leading terms in the Laurent expansions of the
functions $F_{\varphi }^{(E)}\left( \varphi \right) $ and $V_{(E)}(\varphi )$%
.

Introducing a new scalar field $\phi $ according to the relation%
\begin{equation}
\phi =\phi _{0}\ln \left( \varphi /\varphi _{0}\right) ,  \label{phican}
\end{equation}%
with $\varphi _{0}$ being an integration constant and%
\begin{equation}
\phi _{0}=m_{P}^{(D-1)/2}\sqrt{\frac{2D}{D-1}},  \label{phi0}
\end{equation}%
the kinetic term for the scalar field is written in the standard canonical
form:
\begin{equation}
S=\int d^{D}x\sqrt{\left\vert g_{(E)}\right\vert }\left[ -m_{P}^{D-1}R_{(E)}+%
\frac{1}{2}g_{(E)}^{ik}\partial _{i}\phi \partial _{k}\phi -V_{E}(\phi
)+L_{m}^{E}\left( \phi ,g_{(E)ik},\psi \right) \right] .  \label{SEc}
\end{equation}%
Here, the non-gravitational Lagrangian density is expressed as%
\begin{equation}
L_{m}^{E}\left( \phi ,g_{(E)ik},\psi \right) =L_{m}^{(E)}(\varphi
_{0}e^{\phi /\phi _{0}},g_{(E)ik},\psi ).  \label{LmE2}
\end{equation}%
Note that in the Einstein representation we have a direct interaction
between the non-gravitational matter and the scalar field.

In what follows it is convenient to take the integration constant in (\ref%
{phican}) $\varphi _{0}=m_{P}^{D-1}$. With this choice, the Einstein frame
potential in terms of the canonical scalar field takes the form%
\begin{equation}
V_{E}(\phi )=-\exp \left( -\frac{D+1}{D-1}\frac{\phi }{\phi _{0}}\right) %
\left[ f(\eta (\varphi ))+\varphi \eta (\varphi )\right] ,  \label{VEphi}
\end{equation}%
where%
\begin{equation}
\varphi =m_{P}^{D-1}e^{\phi /\phi _{0}},  \label{phi}
\end{equation}%
and the function $\eta (\varphi )$ is obtained by inverting of (\ref{phidef}%
). In the qualitative analysis described below we need also to have the
first and second derivatives of the potential. From (\ref{VEphi}) we can
obtain the expressions%
\begin{equation}
\frac{dV_{E}(\phi )}{d(\phi /\phi _{0})}=\exp \left( -\frac{D+1}{D-1}\frac{%
\phi }{\phi _{0}}\right) f(\eta (\varphi ))-\frac{2}{D-1}V_{E}(\phi ),
\label{VEder1}
\end{equation}%
for the first derivative and%
\begin{equation}
\frac{d^{2}V_{E}(\phi )}{d(\phi /\phi _{0})^{2}}=\exp \left( -\frac{D+1}{D-1}%
\frac{\phi }{\phi _{0}}\right) \left[ \frac{\varphi ^{2}}{f^{\prime \prime
}(\eta (\varphi ))}-\frac{D+3}{D-1}f(\eta (\varphi ))\right] +\frac{%
4V_{E}(\phi )}{(D-1)^{2}},  \label{VEder2}
\end{equation}%
for the second derivative.

Let us consider the form of the potential $V_{E}(\phi )$ for some examples
of the function $f(R)$. A number of specific choices for this function have
been discussed in the literature. In the models with quantum corrections to
the Einstein-Hilbert Lagrangian the function $f(R)$ is of the polynomial
form. A similar structure is obtained in the string-inspired models with the
effective action expanded in powers of the string tension. However, it
should be noted that in both these types of models coming from high-energy
physics, the Lagrangian density in addition to the scalar curvature contains
other scalars constructed from the Riemann tensor. In this context, the $f(R)
$ theories can be considered as models simple enough to be easy to handle
from which we gain some insight in modifications of gravity. In some models
proposed for dark energy the function $f(R)$ contains terms with the inverse
power of the Ricci scalar. For one of the first models of this type $%
f(R)=m_{P}^{D-1}(-R+\gamma /R^{m})$ with $\gamma $ and $m>0$ being constants
\cite{Capo02}. However, there is a matter instability problem in these
models. The model with an additional term $\beta R^{2}$ in the brackets has
been discussed in \cite{Broo06}. Models containing in $f(R)$ exponential
functions of the form $e^{\gamma R}$ and providing the accelerating
cosmological solutions without a future singularity are considered in \cite%
{Cogn08}. Examples of the $f(R)$ functions, containing combinations of the
powers and exponentials of $R$, that allow to construct models with a
late-time accelerated expansion consistent with local gravity
constraints, are studied in references \cite{Hu07} (see also \cite{Noji07}-%
\cite{Clif12}). For example, in the Tsujikawa model $f(R)=m_{P}^{D-1}[-R+%
\gamma \tanh (R/R_{0})]$, whereas in the Hu and Sawicki model $%
f(R)=m_{P}^{D-1}[-R+\gamma (1+(R/R_{0})^{-m})]$ with constants $\gamma $ and
$R_{0}$.

\begin{figure}[tbph]
\begin{center}
\begin{tabular}{cc}
\epsfig{figure=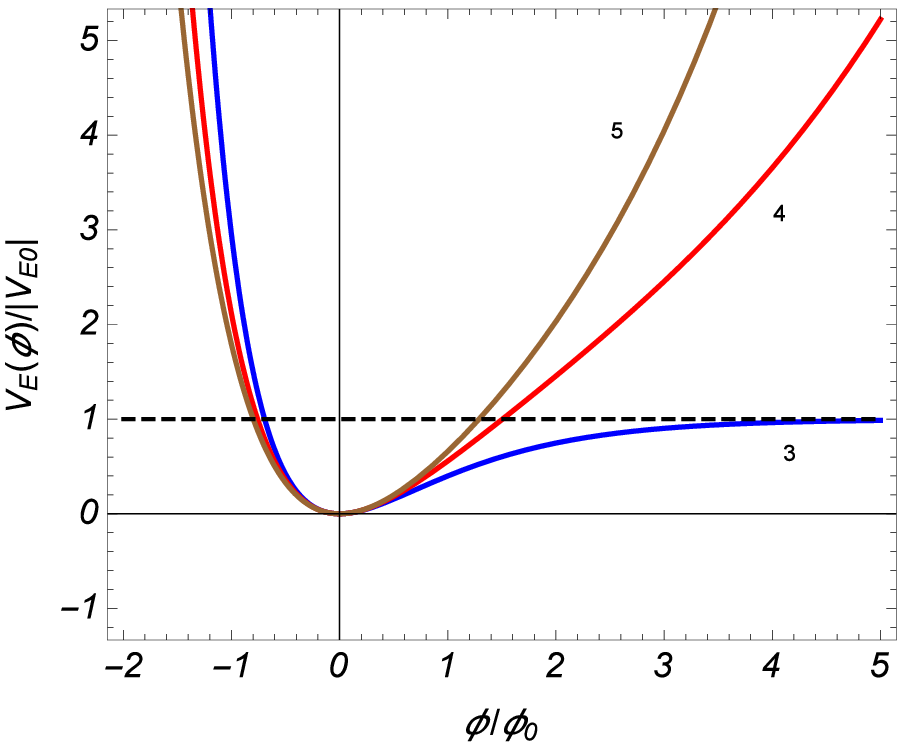,width=6.cm,height=6cm} & \quad %
\epsfig{figure=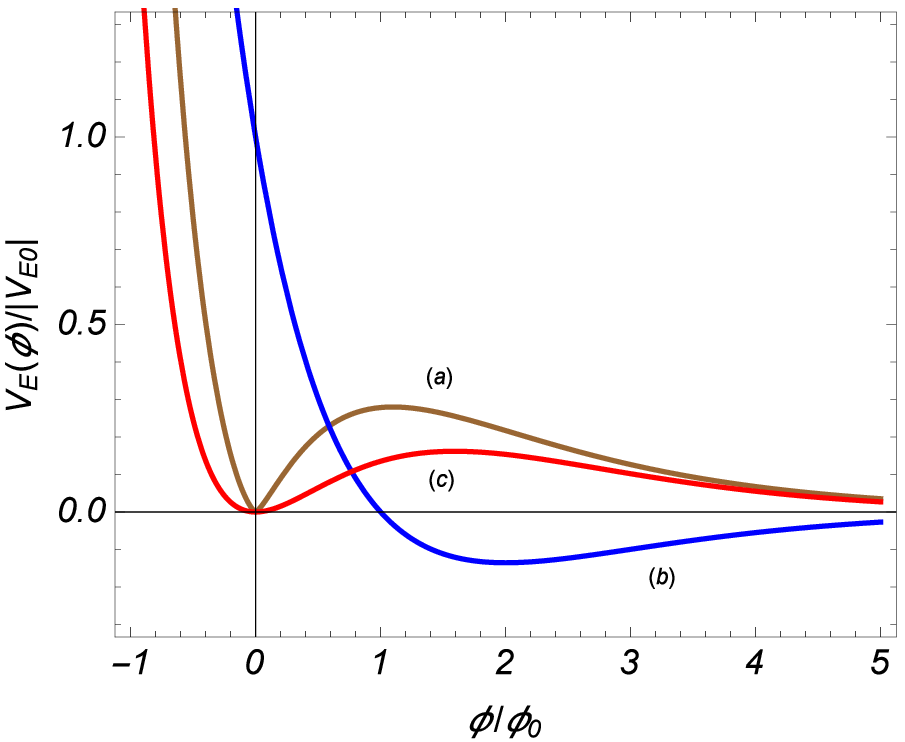,width=6.25cm,height=5.75cm}%
\end{tabular}%
\end{center}
\caption{The left panel presents the Einstein frame potentials in the
Starobinsky model for different values of the spatial dimension (numbers
near the curves). On the right panel the potentials corresponding to the $%
f(R)$ functions (\protect\ref{fRn}) (for $n=4$, curve (a)), (\protect\ref%
{fexp1}) (curve (b)) and (\protect\ref{fexp2}) (curve (c)) are plotted.}
\label{fig1}
\end{figure}

We start our discussion with a $(D+1)$-dimensional generalization of the
Starobinsky model (see \cite{Maed88} for the discussion of inflation in this
type of models). The corresponding lagrangian density for the gravitational
field is taken as%
\begin{equation}
f(R)=m_{P}^{D-1}\left( -R+\beta R^{2}\right) ,  \label{fRStar}
\end{equation}%
where $\beta $ is a constant. The potential in terms of the canonical scalar
field is written in the form%
\begin{equation}
V_{E}(\phi )=V_{E0}\exp \left( \frac{D-3}{D-1}\frac{\phi }{\phi _{0}}\right)
\left( 1-e^{-\phi /\phi _{0}}\right) ^{2},  \label{VEphi2}
\end{equation}%
where $V_{E0}=m_{P}^{D-1}/(4\beta )$. For $\beta >0$ and $\phi \neq 0$ the
potential (\ref{VEphi2}) is positive. It has a minimum at $\phi =0$ with $%
V_{E}(0)=0$. In figure \ref{fig1} we have plotted the potential (\ref{VEphi2}%
) as a function of $\phi /\phi _{0}$ for various values of the spatial
dimension (numbers near the curves). As is seen from the graphs, in the case
$D=3$ an inflationary plateau appears for large values of $\phi /\phi _{0}$
which corresponds to the Starobinsky inflation (for a recent discussion of
the universality of the inflation in the Starobinsky model and its
generalizations see \cite{Kallo13}). Hence, from the point of view of the
Starobinsky inflation, the spatial dimension $D=3$ is special.

As the next example, consider the model with a polynomial function%
\begin{equation}
f(R)=m_{P}^{D-1}\left( -R+\sum_{l=2}^{n}\beta _{l}R^{l}\right) ,
\label{fPol1}
\end{equation}%
with $n\geqslant 2$. For the Einstein frame potential one gets the expression%
\begin{equation}
V_{E}(\phi )=m_{P}^{D-1}\exp \left( -\frac{D+1}{D-1}\frac{\phi }{\phi _{0}}%
\right) \sum_{l=2}^{n}(l-1)\beta _{l}\chi ^{l}(\phi ),  \label{VEPol}
\end{equation}%
where the function $\chi (\phi )$ is defined by the relation%
\begin{equation}
\sum_{l=2}^{n}l\beta _{l}\chi ^{l-1}(\phi )=1-e^{\phi /\phi _{0}}.
\label{RelPol}
\end{equation}%
The relations (\ref{VEPol}) and (\ref{RelPol}) define the potential $%
V_{E}(\phi )$ in the parametric form with $\chi $ being the parameter.

Let us investigate the asymptotics of the potential (\ref{VEPol}) in the
limits $\phi \rightarrow \pm \infty $. From (\ref{RelPol}) it follows that
in the limit $\phi \rightarrow +\infty $ one has $\chi (\phi )\approx
\lbrack -e^{\phi /\phi _{0}}/(n\beta _{n})]^{1/(n-1)}$. In particular, we
see that for an odd $n$ one should have $\beta _{n}<0$. For the asymptotic
behavior of the potential we get%
\begin{equation}
V_{E}(\phi )\approx -\frac{\left( n-1\right) m_{P}^{D-1}}{n^{n/(n-1)}\left(
-\beta _{n}\right) ^{1/(n-1)}}\exp \left[ \left( \frac{1}{n-1}-\frac{2}{D-1}%
\right) \frac{\phi }{\phi _{0}}\right] .  \label{VEinf1}
\end{equation}%
Hence, in the limit $\phi \rightarrow +\infty $ one has $V_{E}(\phi
)\rightarrow 0$ for $n>(D+1)/2$. For $n=(D+1)/2$ the potential tends to the
finite limiting value determined by the coefficient of the exponent in (\ref%
{VEinf1}). In the case $n<(D+1)/2$, the potential tends to $+\infty $ or $%
-\infty $ depending on the sign of the coefficient $\beta _{n}$. In the
limit $\phi \rightarrow -\infty $ the function $\chi (\phi )$ tends to the
finite limiting value $\chi _{-}\equiv \chi (-\infty )$ determined by the
relation $\sum_{l=2}^{n}l\beta _{l}\chi _{-}^{l-1}=1$ (see (\ref{RelPol})).
As a result, the potential behaves as%
\begin{equation}
V_{E}(\phi )\approx m_{P}^{D-1}\exp \left( -\frac{D+1}{D-1}\frac{\phi }{\phi
_{0}}\right) \sum_{l=2}^{n}(l-1)\beta _{l}\chi _{-}^{l},  \label{VEinf2}
\end{equation}%
for $\phi \rightarrow -\infty $. As is seen, the functional form of the
potential in this region is universal and the information on the
coefficients of the polynomial function (\ref{fPol1}) is contained in the
coefficient only.

As a special case of (\ref{fPol1}), let us consider the model%
\begin{equation}
f(R)=m_{P}^{D-1}\left( -R+\beta _{n}R^{n}\right) ,  \label{fRn}
\end{equation}%
with even $n$ and $\beta _{n}>0$. The corresponding potential is nonnegative
and is given by the expression:%
\begin{equation}
V_{E}(\phi )=V_{E0}\exp \left( -\frac{D+1}{D-1}\frac{\phi }{\phi _{0}}%
\right) [(1-e^{\phi /\phi _{0}})^{n}]^{1/(n-1)},  \label{VEpoln}
\end{equation}%
where%
\begin{equation}
V_{E0}=\frac{(n-1)m_{P}^{D-1}}{n^{n/(n-1)}\beta _{n}^{1/(n-1)}}.
\label{V0pol}
\end{equation}%
For $n=2$ the potential is reduced to the one for the Starobinsky model. In
the limit $\phi /\phi _{0}\gg 1$ the potential behaves as
\begin{equation}
V_{E}(\phi )\propto \exp \left[ \left( \frac{1}{n-1}-\frac{2}{D-1}\right)
\frac{\phi }{\phi _{0}}\right] .  \label{VEpolnAs}
\end{equation}%
Hence, in this limit one has $V_{E}(\phi )\rightarrow +\infty $ for $%
n<(D+1)/2$ and $V_{E}(\phi )\rightarrow 0$ for $n>(D+1)/2$. For $n=(D+1)/2$,
in the limit $\phi /\phi _{0}\rightarrow +\infty $ the potential has a
nonzero plateau: $V_{E}(\phi )\rightarrow V_{E0}$. The potential (\ref%
{VEpoln}) for $n=4$ and $D=3$ is depicted in the right panel of figure \ref%
{fig1} (graph (a)).

For the next example we take the function%
\begin{equation}
f(R)=f_{0}e^{\gamma R}.  \label{fexp1}
\end{equation}
The corresponding potential takes the form%
\begin{equation}
V_{E}(\phi )=V_{E0}\exp \left( -\frac{2}{D-1}\frac{\phi }{\phi _{0}}\right) %
\left[ \phi /\phi _{0}+\ln (-M_{D+1}^{D-1}/\gamma f_{0})-1\right] ,
\label{VEexp}
\end{equation}%
with $V_{E0}=-m_{P}^{D-1}/\gamma $. For $|\gamma R|\ll 1$ one has $%
f(R)=f_{0}+f_{0}\gamma R$. Taking $f_{0}\gamma =-m_{P}^{D-1}$, the linear in
$R$ term coincides with the Hilbert-Einstein lagrangian density. With this
choice the potential simplifies to%
\begin{equation}
V_{E}(\phi )=V_{E0}\exp \left( -\frac{2}{D-1}\frac{\phi }{\phi _{0}}\right)
\left( \phi /\phi _{0}-1\right) ,  \label{VEexp2}
\end{equation}%
The graph of this potential for $\gamma >0$ is plotted in the right panel of
figure \ref{fig1} (curve (b)). The value of the potential at the minimum is
negative. Cosmological consequences of this feature will be discussed below.
Note that in this case $f_{0}=-V_{E0}>0$ and for $|\gamma R|\ll 1$ the model
reduces to General Relativity with a negative cosmological constant $V_{E0}$.

In the case of the function%
\begin{equation}
f(R)=f_{0}\left( e^{\gamma R}-1\right) ,  \label{fexp2}
\end{equation}
with $f_{0}\gamma =-m_{P}^{D-1}$ and for small curvatures, corresponding to $%
|\gamma R|\ll 1$, the model is reduced to General Relativity with zero
cosmological constant. The corresponding potential is given by the expression%
\begin{equation}
V_{E}(\phi )=V_{E0}\exp \left( -\frac{D+1}{D-1}\frac{\phi }{\phi _{0}}%
\right) \left[ 1+e^{\phi /\phi _{0}}\left( \phi /\phi _{0}-1\right) \right] ,
\label{VEexp3}
\end{equation}%
with the same notation $V_{E0}$ as in (\ref{VEexp2}). This potential for $%
\gamma <0$ ($V_{E0}>0$) is plotted in figure \ref{fig1} (curve (c)).

\section{Cosmological model}

\label{sec:Cosm}

In this section we consider a homogeneous and isotropic cosmological model
described by the Einstein frame action (\ref{SEc}). The corresponding line
element has the form
\begin{equation}
ds_{E}^{2}=dt^{2}-a^{2}(t)dl^{2}  \label{cosmetric}
\end{equation}%
where $dl$ is the line element of a $D$ - dimensional space of constant
curvature, $a(t)$ is the scale factor. From the homogeneity of the model it
follows that the scalar field should also depend on time only, $\phi =\phi
(t)$. From the field equations we obtain that the energy-momentum tensor
corresponding to the metric (\ref{cosmetric}) is diagonal and can be
presented in the perfect fluid form $T_{i}^{k}=\mathrm{diag}(\varepsilon
,...,-p,...)$, where $\varepsilon $ is the energy density and $p$ is the
effective pressure.

For a model with a flat space, the Einstein frame evolution equations for
the scale factor and the scalar field can be written as%
\begin{eqnarray}
\dot{H}+DH^{2} &=&\frac{m_{P}^{1-D}}{D-1}\left[ \frac{1-w}{2}\varepsilon
+V_{E}\left( \phi \right) \right] ,  \notag \\
\ddot{\phi }+DH\dot{\phi } &=&\alpha \varepsilon -V_{E}^{\prime }\left( \phi
\right) ,  \notag \\
D(D-1)H^{2} &=&m_{P}^{1-D}\left[ \varepsilon +\dot{\phi }^{2}/2+V_{E}\left(
\phi \right) \right] ,  \label{CosmEq2}
\end{eqnarray}%
where the overdot denotes the time derivative and the following notations
are introduced
\begin{equation}
H=\frac{\dot{a}}{a},\;w=\frac{p}{\varepsilon },\;\alpha =\frac{1}{%
\varepsilon \sqrt{|g_{(E)}|}}\frac{\delta L_{m}^{E}\sqrt{|g_{(E)}|}}{\delta
\phi }.  \label{H}
\end{equation}

Excluding $H$ by using the last equation of (\ref{CosmEq2}) and introducing
dimensionless quantities $x=\phi /\phi _{0}$, $\tau =t/t_{0}$, with $t_{0}$
being a positive constant with the dimension of time, for expanding models
the set of cosmological equations is written in terms of the third order
autonomous dynamical system%
\begin{eqnarray}
\frac{dx}{d\tau } &=&y,  \notag \\
\frac{dy}{d\tau } &=&-by[2\epsilon +y^{2}+2V(x)]^{1/2}+\phi _{0}\alpha
\epsilon -V^{\prime }(x),  \label{DynSys} \\
\frac{d\epsilon }{d\tau } &=&-\{b(1+w)[2\epsilon +y^{2}+2V(x)]^{1/2}+\phi
_{0}\alpha y\}\epsilon .  \notag
\end{eqnarray}%
Here we have defined dimensionless functions%
\begin{equation}
V(x)=\left( t_{0}/\phi _{0}\right) ^{2}V_{E}(\phi _{0}x),\;\epsilon
=(t_{0}/\phi _{0})^{2}\varepsilon ,  \label{Vx}
\end{equation}%
and%
\begin{equation}
b=\frac{D}{D-1}.  \label{b}
\end{equation}%
Note that the function $\phi _{0}\alpha $ is dimensionless as well. The
Einstein frame Hubble function is expressed in terms of the variables of the
dynamical system (\ref{DynSys}) as%
\begin{equation}
H^{2}=\frac{2\epsilon +y^{2}+2V(x)}{(D-1)^{2}t_{0}^{2}}.  \label{HE}
\end{equation}

The set of equations (\ref{DynSys}) describes the cosmological dynamics in
the Einstein frame. The corresponding dynamics in the Jordan frame is
obtained by using the conformal transformation (\ref{Conf}) with the
function (\ref{OmE}). For the line element in the Jordan frame one has $%
ds_{J}^{2}=dt_{J}^{2}-a_{J}^{2}(t_{J})dl^{2}$, where the comoving time
coordinate and the scale factor are related to the corresponding Einstein
frame quantities by%
\begin{equation}
dt_{J}=m_{P}\varphi ^{1/(1-D)}dt,\;a_{J}(t_{J})=m_{P}\varphi ^{1/(1-D)}a(t).
\label{RelEJ}
\end{equation}%
From here we get the relation between the Hubble functions in the Einstein
and Jordan frames:%
\begin{equation}
H_{J}(t_{J})=\frac{1}{a_{J}(t_{J})}\frac{da_{J}(t_{J})}{dt_{j}}=\frac{%
\varphi ^{1/(D-1)}}{m_{P}}\left[ H(t)-\frac{\dot{\phi}/\phi _{0}}{D-1}\right]
.  \label{HJrel}
\end{equation}%
Substituting the expression for $H(t)$ from the last equation of (\ref%
{CosmEq2}), this gives%
\begin{equation}
H_{J}=\frac{\varphi ^{1/(D-1)}/\phi _{0}}{\left( D-1\right) m_{P}}\left[ \pm
\sqrt{2\varepsilon +\dot{\phi}^{2}+2V_{E}\left( \phi \right) }-\dot{\phi}%
\right] ,  \label{Hjrel2}
\end{equation}%
where the upper/lower sign corresponds to expanding/contracting models in
the Einstein frame. From the relation (\ref{Hjrel2}) it follows that for $%
V_{E}\left( \phi \right) +\varepsilon >0$ the expansion/contraction in the
Einstein frame corresponds to the expansion/contraction in the Jordan frame.

\section{Qualitative analysis of gravi-scalar models}

\label{sec:Qualit}

The dynamical system (\ref{DynSys}) has an invariant phase plane $%
\varepsilon =0$ which corresponds to the pure gravi-scalar models. First we
consider the qualitative dynamics of these models (for applications of the
qualitative theory of dynamical systems in cosmology see \cite{Wain97}).

\subsection{General analysis}

In what follows it is convenient to introduce dimensionless quantities $%
x=\phi /\phi _{0}$, $\tau =t/t_{0}$, where $t_{0}$ is a positive constant
with the dimension of time. In terms of these variables, for pure
gravi-scalar models the system (\ref{DynSys}) is reduced to the following
second order dynamical system%
\begin{eqnarray}
\frac{dx}{d\tau } &=&y,  \notag \\
\frac{dy}{d\tau } &=&-by[y^{2}+2V(x)]^{1/2}-V^{\prime }(x),  \label{DynSys1}
\end{eqnarray}%
where $\dot{\phi }=\left( \phi _{0}/t_{0}\right) y$ and%
\begin{equation}
b=c\phi _{0}=\frac{D}{D-1},\;V(x)=\left( t_{0}/\phi _{0}\right)
^{2}V_{E}(\phi _{0}x).  \label{bV}
\end{equation}%
For Einstein frame expanding models, the Hubble function is expressed in
terms of the solution of dynamical system (\ref{DynSys1}) as
\begin{equation}
H=\frac{\sqrt{y^{2}+2V(x)}}{(D-1)t_{0}}.  \label{Hubb}
\end{equation}

For nonnegative potentials, introducing the function $X(x)$ in accordance
with the relation $y=\sqrt{2V(x)}\sinh X(x)$, the equation for the phase
trajectories is written as%
\begin{equation}
X^{\prime }(x)=-b-\frac{V^{\prime }(x)}{2V(x)}\coth X(x),  \label{EqPhTraj}
\end{equation}%
This equation is exactly solvable in a special case of exponential
potentials (for a recent discussion of scalar cosmologies with exponential
potentials see \cite{Fre13}):%
\begin{equation}
V(x)=V_{1}e^{\sigma x},  \label{ExpPot}
\end{equation}%
with $V_{1}$ and $\sigma $ being constants. The equation of the phase
trajectories is written in the parametric form as%
\begin{eqnarray}
x &=&\frac{1}{2b}\left[ \frac{2q}{1-q^{2}}\ln |1+qz|+\frac{\ln |z-1|}{1+q}+%
\frac{\ln |z+1|}{q-1}\right] +C,  \notag \\
y &=&\sqrt{2V_{1}}e^{\sigma x/2}\frac{\mathrm{sgn}(z)}{\sqrt{z^{2}-1}},
\label{xyPar}
\end{eqnarray}%
where $q=\sigma /(2b)$, $z=\coth X$ and $C$ is an integration constant. The
corresponding Hubble function is found from (\ref{Hubb}):%
\begin{equation}
H=\frac{\sqrt{2V_{1}}e^{\sigma x/2}|z|}{(D-1)t_{0}\sqrt{z^{2}-1}}.
\label{HubbExp}
\end{equation}%
It can be seen that the limit $z^{2}\rightarrow 1$ corresponds to the early
stages of the cosmological expansion ($\tau \rightarrow 0$). In this limit
one has $y^{2}\gg V(x)$ and the cosmological dynamics is dominated by the
kinetic energy of the scalar field. Under the condition $|\sigma |<2b$, the
limit $z\rightarrow -1/q$ corresponds to the late stages of the expansion, $%
\tau \rightarrow +\infty $. In this limit the kinetic and potential energies
of the scalar field are of the same order: $y^{2}\approx 2V(x)/(q^{-2}-1)$.

For $|\sigma |<2b$ the equation (\ref{EqPhTraj}) has a special solution $%
\coth X=-1/q$. The corresponding phase trajectory is described by the
equation%
\begin{equation}
y=-\frac{\sigma \sqrt{2V_{1}}e^{\sigma x/2}}{\sqrt{4b^{2}-\sigma ^{2}}}.
\label{ySpecial}
\end{equation}%
Note that for this solution the ratio of the kinetic and potential energies
of the scalar field is a constant. For the ratio of the corresponding
pressure and energy density one gets $p_{\phi }/\varepsilon _{\phi }=\sigma
^{2}/(2b^{2})-1$. The time dependence of the special solution is given by%
\begin{eqnarray}
x &=&x_{1}-\frac{2}{\sigma }\ln \tau ,  \notag \\
a(t) &=&\mathrm{const}\,(t/t_{0})^{\beta _{E}},\;\beta _{E}=\frac{4D}{%
(D-1)^{2}\sigma ^{2}},  \label{atSpec}
\end{eqnarray}%
where%
\begin{equation}
x_{1}=\left( 2\frac{4b^{2}/\sigma ^{2}-1}{\sigma ^{2}B}\right) ^{1/\sigma }.
\label{x1}
\end{equation}%
For $|\sigma |<2\sqrt{D}/(D-1)$, the expansion described by (\ref{atSpec})
corresponds to a power-law inflation in the Einstein frame. The special
solution (\ref{atSpec}) is a future attractor ($t\rightarrow +\infty $) for
a general solution (\ref{xyPar}).

Now we turn to the qualitative analysis of the system (\ref{DynSys1}) for
the general case of the potential $V(x)$ (for applications of the
qualitative theory of dynamical systems in cosmology see \cite{Wain97}). The
critical points for the system are the points of the phase space $(x,y)$
with the coordinates $(x_{c},0)$ where $V^{\prime }(x_{c})=0$, $%
V(x_{c})\geqslant 0$. For the corresponding solution the Hubble function is
a constant, $H=H_{c}$, with%
\begin{equation}
H_{c}^{2}=\frac{V_{E}(\phi _{c})}{m_{P}^{D-1}D(D-1)}.  \label{Hc}
\end{equation}%
This solution describes the Minkowski spacetime for $V_{E}(\phi _{c})=0$ and
the de Sitter spacetime for $V_{E}(\phi _{c})>0$. In the latter case for the
cosmological constant one has $\Lambda =V_{E}(\phi _{c})/(2m_{P}^{D-1})$.

The character of the critical points is defined by the eigenvalues%
\begin{equation}
\lambda _{1,2}=-b\sqrt{V_{c}/2}\pm \sqrt{b^{2}V_{c}/2-V_{c}^{\prime \prime }}%
,  \label{lamb}
\end{equation}%
where $V_{c}=V(x_{c})$, $V_{c}^{\prime \prime }=V^{\prime \prime }(x_{c})$.
From here it follows that for $V_{c}^{\prime \prime }<0$ ($x_{c}$ is a
maximum of the potential $V(x)$) the critical point is a saddle. The
directions of the corresponding separatrices are determined by the unit
vectors $\mathbf{n}^{(i)}=(1,\lambda _{i})/\sqrt{1+\lambda _{i}^{2}}$, $%
i=1,2 $. For $V_{c}^{\prime \prime }>0$ ($x_{c}$ is a minimum of the
potential $V(x)$) two cases should be considered separately. When $%
0<V_{c}^{\prime \prime }<b^{2}V_{c}/2$, the critical point is a stable node.
For $0<b^{2}V_{c}/2<V_{c}^{\prime \prime }$ the critical point is a stable
sink. In the case $V_{c}>0$, $V_{c}^{(i)}\equiv (d^{i}V/dx^{i})_{x=x_{c}}=0$
for $i=1,\ldots ,n-1$, and $V_{c}^{(n)}\neq 0$, the critical point is (i) a
saddle for even $n$ and $V_{c}^{(n)}<0$, (ii) a stable node for even $n$ and
$V_{c}^{(n)}>0$, (iii) a degenerate critical point with one stable node
sector and with two saddle sectors for odd $n$. Another degenerate case
corresponds to $V_{c}=0$ and $V_{c}^{\prime \prime }>0$. In this case the
critical point is a stable sink.

By using the expressions (\ref{VEder1}) and (\ref{VEder2}), we can express $%
V_{c}$ and $V_{c}^{\prime \prime }$ in terms of the function $f(R)$:
\begin{eqnarray}
V_{c} &=&\frac{D-1}{2}\exp \left( -\frac{D+1}{D-1}x_{c}\right) f_{c},  \notag
\\
V_{c}^{\prime \prime } &=&\exp \left( \frac{D-3}{D-1}x_{c}\right) \frac{%
m_{P}^{2(D-1)}}{f_{c}^{\prime \prime }}-\frac{2(D+1)}{(D-1)^{2}}V_{c},
\label{Vc}
\end{eqnarray}%
where%
\begin{equation}
f_{c}=f(\eta (\varphi _{c})),\;f_{c}^{\prime \prime }=f^{\prime \prime
}(\eta (\varphi _{c})),  \label{fc}
\end{equation}%
and $\varphi _{c}=m_{P}^{D-1}e^{x_{c}}$. Note that $R_{c}=\eta (\varphi
_{c}) $ corresponds to the Ricci scalar in the original representation (\ref%
{SfR}), evaluated at the critical point. By taking into account that $%
V_{c}\geqslant 0$, from (\ref{Vc}) we conclude that the critical points
correspond to the values of the Ricci scalar for which $f(R_{c})\geqslant 0$%
. We also see that $V_{c}^{\prime \prime }<0$ for $f_{c}^{\prime \prime }<0$
and, hence, in this case the critical point is unstable being a saddle point.

We should also consider the behavior of the phase trajectories at the
infinity of the phase plane. With this aim, it is convenient to introduce
polar coordinates $(\rho ,\theta )$ defined as
\begin{equation}
x=\frac{\rho \cos \theta }{1-\rho },\;y=\frac{\rho \sin \theta }{1-\rho },
\label{Map}
\end{equation}%
with $0\leqslant \rho \leqslant 1$, $0\leqslant \theta \leqslant 2\pi $. Now
the phase space is mapped onto a unite circle. The points at infinity
correspond to $\rho =1$. For the potentials having the asymptotic behavior $%
V(x)\sim B|x|^{m}$, $m<4$, in the limit $x\rightarrow \infty $ one has the
following critical points on the circle $\rho =1$. The points $\theta =0$
and $\theta =\pi $ are stable nodes for $m<0$ and saddles with two sectors
for $m>0$. In the latter case the sectors are separated by a special
solution described by the trajectory
\begin{equation}
y(x)\approx -\frac{V^{\prime }(x)}{b\sqrt{2V(x)}}\sim \frac{m}{b}\sqrt{B/2}%
|x|^{m/2-1},  \label{SpSol}
\end{equation}%
for $|x|\rightarrow \infty $. In the vicinity of the points $\theta =\pi /2$
and $\theta =3\pi /2$ the potential terms can be neglected and these points
are unstable degenerate nodes. For $m=4$ the nature of the critical points
at $\theta =\pi /2$ and $\theta =3\pi /2$ remains the same. In this case the
other critical points correspond to $\theta =-\arctan (\sqrt{8B}/b)$ and $%
\theta =\pi -\arctan (\sqrt{8B}/b)$. The phase portrait near these points
have two saddle sectors which are separated by the trajectory corresponding
to the special solution (\ref{SpSol}). For $m>4$ there are two critical
points on the circle $\rho =1$ corresponding to $\theta =\pi /2$ and $\theta
=3\pi /2$. These points are degenerate and have an unstable node sector and
a saddle sector separated by the special solution (\ref{SpSol}). Similar
behavior of the phase trajectories at the infinity takes place for the
potentials with the asymptotic behavior $V(x)\sim Be^{\sigma |x|}$, $\sigma
>0$, for $x\rightarrow \infty $ and for the values of the parameter $%
0<\sigma <2b$. The separatrix between the saddle and node sectors is
described by the special solution $y\approx -\mathrm{sgn}(x)\sigma \sqrt{2B}%
e^{\sigma |x|/2}/\sqrt{4b^{2}-\sigma ^{2}}$ for $x\rightarrow \infty $. The
general solution behaves as $y\approx -\mathrm{sgn}(x)Ce^{b|x|}$, with a
positive constant $C$. This behavior coincides with that in the absence of
the potential. For $\sigma \geqslant 2b$ the dynamical system (\ref{DynSys1}%
) has no critical points at infinity (on the circle $\rho =1$).

We have described the general evolution of gravi-scalar models. In the
presence of a barotropic non-gravitational matter with $w=\mathrm{const}$
the evolution is described by the three-dimensional dynamical system (\ref%
{DynSys}) with the phase space $(x,y,\epsilon )$. Note that, as a
consequence of the term $\phi _{0}\alpha y$ in the equation for $d\epsilon
/d\tau $, in expanding models the energy density, in general, is not a
monotonically decreasing function of the Einstein frame time coordinate for $%
w>-1$. The phase trajectories corresponding to gravi-scalar models lie in
the plane $(x,y,0)$ that forms an invariant subspace. The points $(x_{c},0,0)
$ of this subspace are critical points of the system (\ref{DynSys1}). For
the corresponding eigenvalues one has $(\lambda _{1},\lambda _{2},\lambda
_{3})$, where $\lambda _{1,2}$ are given by the expression (\ref{lamb}) and $%
\lambda _{3}=-b(1+w)\sqrt{2V_{c}}$. Notice that these eigenvalues do not
depend on the function $\alpha $. We see that for $w>-1$ and $V_{c}>0$ the
critical point $(x_{c},0,0)$ is a stable local attractor for general
cosmological solutions with barotropic matter. The corresponding geometry is
the de Sitter spacetime with the scale factor $a(t)=\mathrm{const}\,\exp [%
\sqrt{2V_{c}}\tau /(D-1)]$. Near the critical point $(x_{c},0,0)$ the energy
density behaves as $\varepsilon \sim \exp [-b(1+w)\sqrt{2V_{c}}\tau ]$, $%
\tau \rightarrow +\infty $, and it decays exponentially. Hence, the late time
evolution of the corresponding models is governed by the effective
cosmological constant determined by the value of the potential at its
minimum. For the phantom matter, $w<-1$, one has $\lambda _{3}>0$ and the
critical point $(x_{c},0,0)$ is unstable. For the corresponding models the
late time dynamics is driven by the non-gravitational matter. The
qualitative analysis is more complicated for $V_{c}=0$. In this case the
critical point is degenerate and in order to determine the behaviour of the
phase trajectories in its neighborhood one needs to keep nonlinear terms in
the expansions of the right-hand sides of (\ref{DynSys1}). The complete
analysis of the models with a barotropic matter, including the points at the
infinity of the phase space, will be discussed elsewhere.

\subsection{Qualitative analysis in special cases}

\begin{figure}[tbph]
\begin{center}
\begin{tabular}{cc}
\epsfig{figure=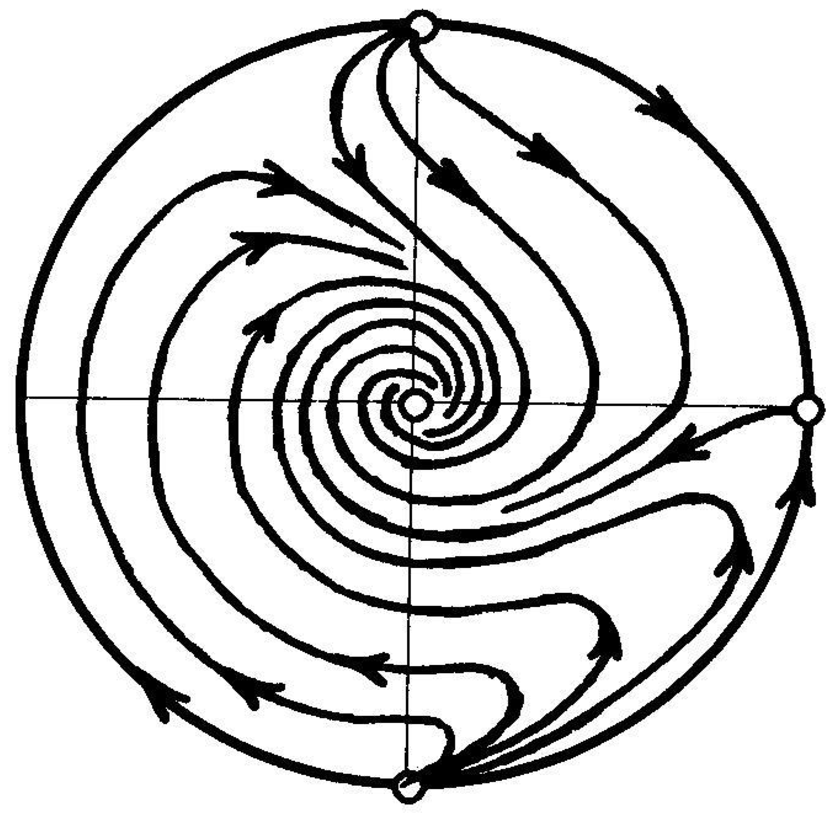,width=6.cm,height=6cm} & \quad %
\epsfig{figure=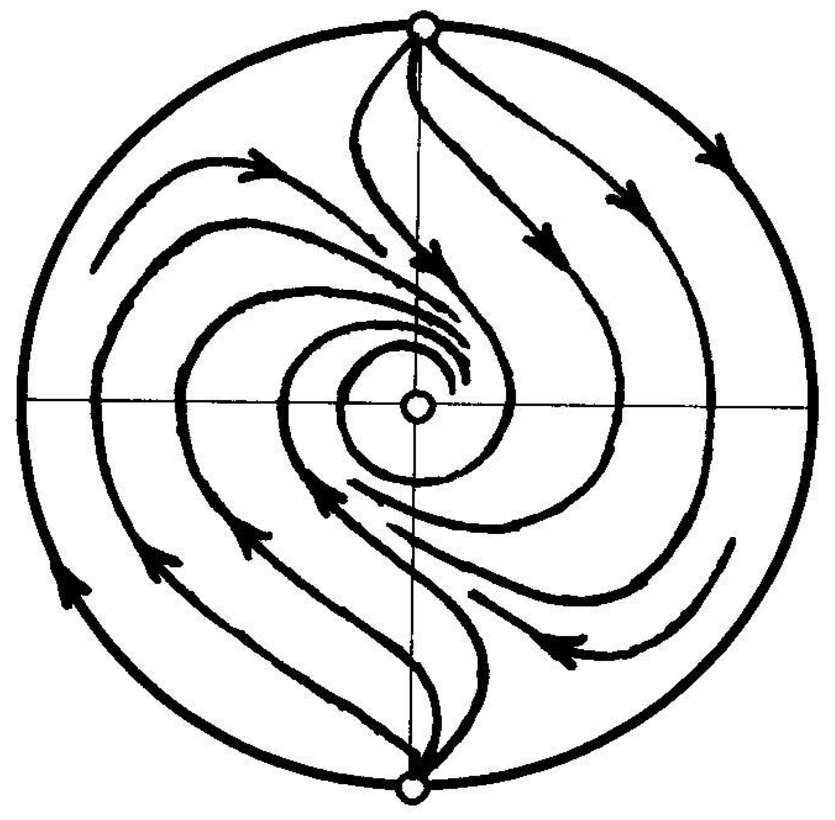,width=6.cm,height=6cm}%
\end{tabular}%
\end{center}
\caption{Phase portraits of the dynamical system for a $(D+1)$-dimensional
generalization of the Starobinsky model in the cases $D=3$ (left panel) and $%
D>3$ (right panel).}
\label{fig2}
\end{figure}

As an application of general analysis given above, first let us consider the
Starobinsky model. The corresponding potential has the form (\ref{VEphi2})
with $\phi /\phi _{0}=x$. In the limit $x\rightarrow -\infty $ the potential
behaves as $\exp [-(D+1)x/(D-1)]$. For the corresponding parameter $\sigma $
one has $\sigma =(D+1)/(D-1)$ and, hence, $\sigma <2b$. From here it follows
that the point $\rho =1$, $\theta =\pi /2$ is degenerate having an unstable
node sector and a saddle sector (see figure \ref{fig2}). In the limit $%
x\rightarrow +\infty $ one has $V(x)\propto \exp [(D-3)x/(D-1)]$ and for $%
D>3 $ the behavior of the phase trajectories near the point $\rho =1$, $%
\theta =3\pi /2$ is similar to that for the point $\rho =1$, $\theta =\pi /2$%
. In the special case $D=3$ the dynamical system has a critical point at $%
\rho =1$, $\theta =0$. This point is a node (see the left panel in figure %
\ref{fig2}) and the corresponding unstable separatarix describes an
inflationary expansion. This special solution is an attractor for the
general solution. For $D=3$ the point $\rho =1$, $\theta =3\pi /2$ at the
infinity of the phase plane is an unstable node. The only critical point in
the finite region of the phase plane, $(x,y)=(0,0)$, corresponds to the
minimum of the potential. This point is a stable sink and the corresponding
geometry is the Minkowski spacetime. The phase portrait, mapped on the unit
circle with the help of (\ref{Map}), is presented in the left panel of
figure \ref{fig2} for $D=3$ and in the right panel for $D>3$.

\begin{figure}[tbph]
\begin{center}
\begin{tabular}{cc}
\epsfig{figure=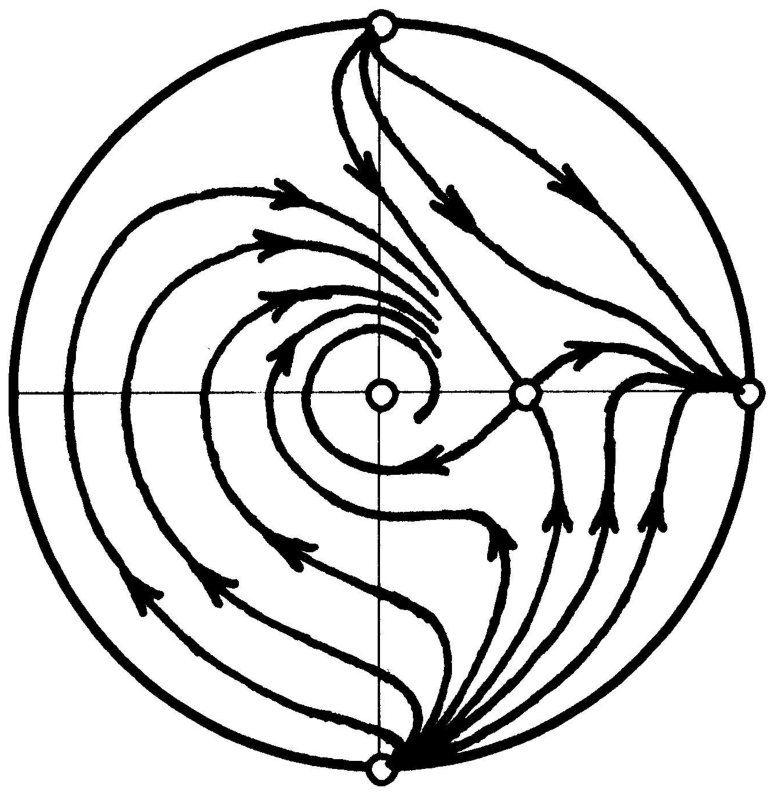,width=6.cm,height=6cm} & \quad %
\epsfig{figure=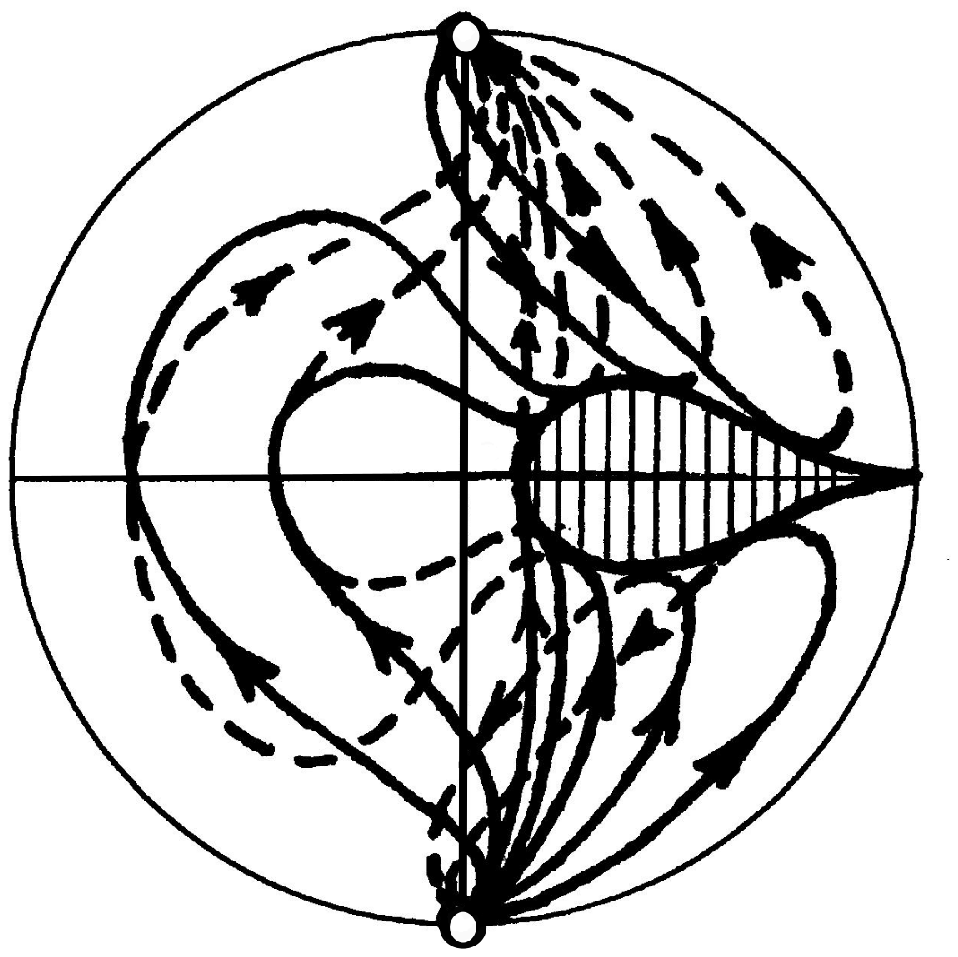,width=6.cm,height=6cm}%
\end{tabular}%
\end{center}
\caption{Phase portraits of the dynamical system for the potential (\protect
\ref{VEpoln}), with $n=4$, and for the potential (\protect\ref{VEexp2}).}
\label{fig3}
\end{figure}

For the model (\ref{fRn}) with an even $n$ and $\beta _{n}>0$, the potential
has the form (\ref{VEpoln}) with $\phi /\phi _{0}=x$. For $n>(D+1)/2$ one
has $V(x)\rightarrow 0$ in the limit $x\rightarrow +\infty $ (see graph (a)
in the right panel of figure \ref{fig1}). In this case we have two critical
points in the finite region of the phase plane. The first one, $(x,y)=(0,0)$%
, corresponds to the minimum of the potential and is a stable sink. The
second one, $(x,y)=(x_{c},0)$, corresponds to the maximum of the potential
and is a saddle. The phase portrait is depicted in the left panel of figure %
\ref{fig3}. At infinity of the phase plane, the nature of the point $\rho =1$%
, $\theta =\pi /2$ remains the same as in the previous example, whereas the
point $\rho =1$, $\theta =3\pi /2$ becomes an unstable node. In the region $%
x\gg 1$, the potential is approximated by an exponential one, (\ref{ExpPot}%
), with $\sigma =1/(n-1)-2/(D-1)$. For $n>(D+1)/2$ one has $\sigma <0$ and
the special solution with the asymptotic behavior (\ref{atSpec}) in the
limit $t\rightarrow +\infty $ is an attractor for a general solution. Note
that for the corresponding value of the parameter $\beta _{E}$ we have
\begin{equation}
\beta _{E}=\left[ \frac{2\sqrt{D}(n-1)}{2n-D-1}\right] ^{2}>1,  \label{bet0}
\end{equation}%
and the solution (\ref{atSpec}) describes a late-time power-law inflation.

On the base of the general analysis in the previous subsection we can also
plot the phase portraits for a general polynomial function (\ref{fPol1}).
The asymptotics of the corresponding potential in the regions $\phi
\rightarrow +\infty $ and $\phi \rightarrow -\infty $ are given by (\ref%
{VEinf1}) and (\ref{VEinf2}), respectively. If the coefficients of the
exponents in these asymptotics are positive, the phase portraits in the
regions $\phi \rightarrow \pm \infty $ are qualitatively equivalent to the
one given on the right panel of figure \ref{fig2} in the case $n<(D+1)/2$,
and to the one on the left panel for $n=(D+1)/2$. In the case $n>(D+1)/2$,
the phase portraits at the infinity of the phase plane is similar to that
plotted on the left panel of figure \ref{fig3}. If the coefficients in the
asymptotic expressions (\ref{VEinf1}) and (\ref{VEinf2}) are negative the
potential goes to $-\infty $ in the limits $\phi \rightarrow \pm \infty $.
In this case the regions of the phase plane near the points $\rho =1$, $%
\theta =0$ and $\rho =1$, $\theta =\pi $ are classically forbidden and one
has the transition from the expanding models to the contracting ones at the
border of the forbidden region, similar to the one described on the right
panel of figure \ref{fig3} (see below). For a general polynomial function (%
\ref{fPol1}), instead of a single minimum, the corresponding potential can
have a set of local minima in the finite region of the phase plane. In this
case, the corresponding phase portrait is divided into regions which are
separated by the stable separatrices of the saddles corresponding to
neighboring local maxima of the potential. If the value of the potential at
the minimum between these maxima is nonnegative, then one has a stable
critical point corresponding to this minimum and it is an attractor (in the
limit $t\rightarrow +\infty $) for all the trajectories between the stable
separatrices of the neighboring saddles. Depending on the value of the
potential at the local minimum the corresponding critical point can be
either a stable sink or a stable node. If the value of the potential at the
minimum is negative, then there is a classically forbidden region in the
phase space $(x,y)$. This region is determined by the inequality%
\begin{equation}
y^{2}+2V(x)<0.  \label{Forb}
\end{equation}%
At the boundary of the forbidden region, given by $y^{2}+2V(x)=0$, one has $%
H=0$ and $\dot{H}=m_{P}^{1-D}V_{E}(\phi )/(D-1)<0$. Hence, at the boundary
the expansion stops at a finite value of the cosmological time $t$ and then
the model enters the stage of the contraction ($H<0$). The corresponding
dynamics is described by the dynamical system (\ref{DynSys1}) with the
opposite sign of the first term in the right-hand side of the second
equation. Note that for nonnegative potentials the expansion-contraction
transition in models with flat space is not classically allowed.

For the function (\ref{fexp2}) with $f_{0}\gamma =-m_{P}^{D-1}$ the
potential is given by the expression (\ref{VEexp3}). In the case $\gamma <0$
the qualitative behavior of this potential is similar to that for the
function (\ref{fRn}) with $n>(D+1)/2$ and the corresponding phase portrait
is qualitatively equivalent to the one presented in the left panel of figure %
\ref{fig3}. However, note that the asymptotic behavior of the potential in
the limit $\phi \rightarrow +\infty $ is not purely exponential. For the
corresponding potential in (\ref{DynSys1}) from (\ref{VEexp3}) one has $%
V(x)\approx V_{2}xe^{-2x/(D-1)}$ in the limit $x\gg 1$. Here, $V_{2}$ is
expressed in terms of the coefficient $V_{E0}$ in (\ref{VEexp3}). It can be
seen that the dynamical system has a special solution with the asymptotic
behavior $y^{2}\approx 2V(x)/(D^{2}-1)$ in the region $x\gg 1$ (compare with
the special solution (\ref{ySpecial}) for the case of pure exponential
potentials). This special solution is an attractor for the general solution
near the critical point $(\rho ,\theta )=(1,0)$. The corresponding time
dependence of the scalar field is determined from the relation $2V(x)\approx
\left( D-1\right) ^{2}(D^{2}-1)/\tau ^{2}$, which is obtained by the
integration of the first equation in (\ref{DynSys1}). With the help of this
relation, the asymptotic behavior of the Einstein frame scale factor is
found from (\ref{Hubb}): $a(t)\approx a_{0}(t/t_{0})^{D}$, $t\rightarrow
+\infty $. The asymptotic behavior of $x$ near the critical point $(\rho
,\theta )=(1,0)$, as a function of the time coordinate is simpler in the
Jordan frame. By using the expression for the function $y(x)$ and the
relation $dt_{J}=e^{-x/(D-1)}dt$, in the region $x\gg 1$ we can see that $%
x\approx V_{2}\tau _{J}^{2}/[2\left( D^{2}-1\right) ]$, where $\tau
_{J}=t_{J}/t_{0}$. For the scale factor in the Jordan frame one gets $%
a_{J}(t_{J})\approx \mathrm{const\,}e^{x}$.

In the case of the function (\ref{fexp1}) with $f_{0}\gamma =-m_{P}^{D-1}$,
the potential is given by the expression (\ref{VEexp2}). For $\gamma >0$ and
in the limit of small curvatures, this model reduces to General Relativity
with a negative cosmological constant. A characteristic feature of the
potential is the presence of the region in the field space where it is
negative. For this type of potentials there is a classically forbidden
region determined by (\ref{Forb}). As it has been noted above, at the
boundary of this region the expansion stops at a finite value of the
cosmological time $t$ and then the model enters the stage of the
contraction. For the potential (\ref{VEexp2}), the only critical points of
the dynamical system (\ref{DynSys1}) are at the infinity of the phase plane.
The corresponding phase portrait is depicted in the right panel of figure %
\ref{fig3}. The classically forbidden region of the phase space is shaded.
The full/dashed trajectories correspond to the expansion/contraction phases.
As it follows from (\ref{DynSys1}), the trajectories for the contraction
stage are obtained from those describing an expansion by the transformation $%
\tau \rightarrow -\tau $, $y\rightarrow -y$. For expanding models, near the
point $\rho =1$, $\theta =\pi /2$ the phase portrait has two sectors: an
unstable node sector and a saddle sector. The point $\rho =1$, $\theta =3\pi
/2$ is an unstable node. Depending on the initial conditions, the expanding
models start their evolution at finite cosmic time $t=t_{i}$ from the point $%
\rho =1$, $\theta =\pi /2$ or from the point $\rho =1$, $\theta =3\pi /2$.
During a finite time interval the trajectories reach the boundary of the
forbidden region (\ref{Forb}) at $t=t_{c}>t_{i}$. At this moment the
expansion stops ($H(t_{c})=0$) and the model enters the contraction stage
(dashed trajectories on the phase portrait). The corresponding trajectories
enter the critical points $\rho =1$, $\theta =3\pi /2$ and $\rho =1$, $%
\theta =\pi /2$ at finite time $t_{f}>t_{c}$. Hence, all the models have a
finite lifetime $t_{f}-t_{i}$.

For the Jordan frame Hubble function, from (\ref{Hjrel2}) for the
gravi-scalar models one gets%
\begin{equation}
H_{J}=\frac{e^{x/(D-1)}}{(D-1)t_{0}}[\pm \sqrt{y^{2}+2V(x)}-y].
\label{HjgrSc}
\end{equation}%
The corresponding comoving time coordinate is determined from the relation $%
dt_{J}=e^{-x/(D-1)}dt$. Similar to the Einstein frame, for nonnegative
potentials, $V(x)\geqslant 0$, the expansion and the contraction models in
the Jordan frame are separated by a classically forbidden region. For
potentials with a region where $V(x)<0$, for the models with the initial
expansion the Jordan frame Hubble function vanishes at the points of the
phase space $(x_{c},y)$, where $x_{c}$ is the zero of the potential $V(x)$, $%
V(x_{c})=0$. The expansion to contraction transition occurs at these points.
The equation of the transition curve on the $(\rho ,\theta )$ plane, defined
in accordance with (\ref{Map}), is given by $\rho =1/[1+\cos (\theta
)/x_{c}] $. For the example presented in the right panel of figure \ref{fig3}%
, compared with the Einstein frame, the expansion to contraction transition
in the Jordan frame occurs earlier (later) for the expansion trajectories
originating from the point $\rho =1$, $\theta =\pi /2$ ($\rho =1$, $\theta
=3\pi /2$).

As an example of comparison of the cosmological dynamics in different frames
let us consider the special solution (\ref{atSpec}) for the exponential
potential (\ref{ExpPot}) in the Jordan frame. In the case $\sigma \neq
-2/(D-1)$, for the corresponding cosmic time one gets
\begin{equation}
t_{J}=\frac{t_{0}\tau ^{u_{0}}}{u_{0}x_{1}^{1/(D-1)}},\;u_{0}\equiv \frac{2}{%
\sigma (D-1)}+1.  \label{tJ}
\end{equation}%
As is seen, we have $0<t_{J}<\infty $ for $u_{0}>0$ and $-\infty <t_{J}<0$
for $u_{0}<0$. For the scalar field and the scale factor in the Jordan frame
we find the expressions%
\begin{eqnarray}
\varphi (t_{J}) &=&m_{P}^{D-1}x_{1}^{b}(u_{0}t_{J}/t_{0})^{-2/(\sigma
u_{0})},  \notag \\
a_{J}(t_{J}) &=&\mathrm{const}\,(u_{0}t_{J}/t_{0})^{\beta _{J}},\;\beta _{J}=%
\frac{1+2b/\sigma }{1+\sigma (D-1)/2}.  \label{SpSolJ}
\end{eqnarray}%
Note that $u_{0}\beta _{J}=(u_{0}-1)^{2}[D+\sigma (D-1)/2]$. By taking into
account that the special solution under consideration is present for $%
\left\vert \sigma \right\vert <2b$, we conclude that $u_{0}\beta _{J}>0$.
Depending on the value of the parameter $\sigma $, we have three
qualitatively different types of evolutions. In the region $-\infty <\sigma
<-2/(D-1)$ one has $u_{0},\beta _{J}>0$, $0<t_{J}<\infty $ and, hence, $%
a_{J}(t_{J})\rightarrow +\infty $, $\varphi (t_{J})\rightarrow +\infty $ for
$t_{J}\rightarrow +\infty $. For $-2/(D-1)<\sigma <0$ we get $u_{0},\beta
_{J}<0$, $-\infty <t_{J}<0$. In this case $a_{J}(t_{J})\rightarrow +\infty $%
, $H_{J}\rightarrow \infty $, $\varphi (t_{J})\rightarrow \infty $ in the
limit $t_{J}\rightarrow 0$ and the point $t_{J}=0$ corresponds to a future
singularity of the "Big Rip" type (for recent discussions of different types
of future cosmological singularities see \cite{Noji05}). For $\sigma >0$ we
have $u_{0},\beta _{J}>0$, $0<t_{J}<\infty $ and $a_{J}(t_{J})\rightarrow
+\infty $, $\varphi (t_{J})\rightarrow 0$ in the limit $t_{J}\rightarrow
+\infty $.

For $\sigma =-2/(D-1)$ one has the relation%
\begin{equation}
t_{J}=t_{0}\frac{D(D-1)}{\sqrt{2V_{1}}}\ln \tau ,  \label{tJs}
\end{equation}%
with $-\infty <t_{J}<+\infty $, and the special solution for the exponential
potential (\ref{ExpPot}) takes the form%
\begin{eqnarray}
\varphi (t_{J}) &=&m_{P}^{D-1}x_{1}\exp \left( \frac{\sqrt{2V_{1}}}{Dt_{0}}%
t_{J}\right) ,  \notag \\
a_{J}(t_{J}) &=&\mathrm{const}\,\exp \left( \frac{\sqrt{2V_{1}}}{Dt_{0}}%
t_{J}\right) .  \label{SpSolJs}
\end{eqnarray}%
In this case we have an exponential inflation in the Jordan frame.

For the model (\ref{fRn}), the potential in the region $x\gg 1$ is
approximated by an exponential one with $\sigma =1/(n-1)-2/(D-1)$. In this
case for the Jordan frame parameters in (\ref{SpSolJ}) one has
\begin{equation}
u_{0}=\frac{D-1}{D+1-2n},\;\beta _{J}=\frac{2(n-1)\left( 2n-1\right) }{D+1-2n%
}.  \label{u0}
\end{equation}%
For $n>(D+1)/2$ these parameters are negative and, hence, $-\infty <t_{J}<0$%
. In this case we have a "Big Rip" type singularity at $t_{J}=0$ in the
Jordan frame. This is the case for the example presented in the left panel
of figure \ref{fig3}. Note that, unlike to the Einstein frame, the
trajectories of the general solution enter the critical point $(\rho ,\theta
)=(1,0)$ at finite value of the Jordan frame time coordinate $t_{J}$.

\section{Conclusion}

\label{sec:Conc}

In the present paper we have considered the qualitative evolution of
cosmological models in $(D+1)$-dimensional $f(R)$ gravity. In order to do
that the model is transformed to an equivalent model described by a
scalar-tensor theory with the action (\ref{SfRc}). By a conformal
transformation one can present the theory in various representations. From
the point of view of the description of the cosmological dynamics, the most
convenient representation corresponds to the Einstein frame, in which the
gravitational part of the action coincides with that for General Relativity.
In this frame there is a direct interaction of the scalar field with a
non-gravitational matter.

For homogeneous and isotropic cosmological models with flat space the
dynamics is described by the set of equations (\ref{CosmEq2}). These
equations can be presented in the form of a third order autonomous dynamical
system (\ref{DynSys}). The corresponding phase space has an invariant
subspace describing the gravi-scalar models in the absence of a
non-gravitational matter. The dynamical system for these models is presented
in the form (\ref{DynSys1}). For a general case of the function $f(R)$, we
have found the critical points of the system and their nature, including the
points at the infinity of the phase plane. As applications of general
analysis, various special cases of the function $f(R)$ are considered. As
the first example, we have taken the $(D+1)$-dimensional generalization of
the Starobinsky model with a quadratic function $f(R)$. For $D=3$ the
corresponding phase portrait is depicted in the left panel of figure \ref%
{fig2}. In this case, the potential has a plateau in the limit $\phi
\rightarrow +\infty $ which describes an inflationary expansion. The
corresponding special solution, presented by the separatrix of the saddle
point in the phase portrait, is an attractor of the general solution. The
latter feature shows that the inflation is a general feature in these
models. From the point of view of the Starobinsky inflation, the spatial
dimension $D=3$ is special: for $D>3$ the inflationary attractor
corresponding to the plateau of the potential is absent and the only stable
critical point corresponds to the minimum of the potential (left panel of
figure \ref{fig2}). The solution corresponding to the latter is Minkowski
spacetime.

We have also considered a polynomial generalization of the Starobinsky
model. In particular, in the model given by (\ref{fRn}) with even $n$, the
inflationary plateau is realized for $n=(D+1)/2$. For $n>(D+1)/2$ and in the
limit $\phi \rightarrow +\infty $ the potential decays exponentially. In
this region models with power-law inflation are realized. The corresponding
phase portrait for $n=4$ is depicted in the left panel of figure \ref{fig3}.
Depending on the initial conditions two classes of cosmological models are
realized. For the first one, presented by the phase trajectories on the left
of the stable separatrices of the saddle point, corresponding to the maximum
of the potential, the spacetime geometry tends to the Minkowski one in the
limit $t\rightarrow +\infty $. For the models from the second class the
future attractor is at infinity of the phase space (the critical point $%
(\rho ,\theta )=(1,0)$). The asymptotic behavior of the scale factor for $%
t\rightarrow +\infty $ is given by (\ref{atSpec}) with the power (\ref{bet0}%
). It presents a power-law inflation in the Einstein frame. In the Jordan
frame the corresponding asymptotic is given by (\ref{SpSolJ}) with the
parameters (\ref{u0}) and describes a "Big Rip" type of singularity.

As further applications of general procedure, we have discussed two examples
of the exponential function $f(R)$. For the first one $f(R)=-m_{P}^{D-1}e^{%
\gamma R}/\gamma $, $\gamma >0$, and in the weak field limit the model
reduces to General Relativity with a negative cosmological constant. The
potential is given by (\ref{VEexp2}) and the phase portrait is presented in
the right panel of figure \ref{fig3}. The shaded region corresponds to the
classically forbidden region in the phase plane. This type of regions arise
for potentials taking negative values in some range of the field space. At
the boundary of the forbidden region the expansion stops and the model
enters the contraction phase. For the second example $f(R)=-m_{P}^{D-1}%
\left( e^{\gamma R}-1\right) /\gamma $, $\gamma <0$, and the potential is
given by (\ref{VEexp3}). This potential is non-negative everywhere and has a
minimum with a zero cosmological constant. The corresponding phase portrait
is qualitatively equivalent to the one depicted in the left panel of figure %
\ref{fig3}. In this example the asymptotic behavior of the potential in the
limit $\phi \rightarrow +\infty $ is not purely exponential. In this region
one has a power-law inflation in the Einstein frame and the exponential
inflation in the Jordan frame.

\section{Acknowledgments}

R. M. A. and G. H. H. were supported by the State Committee of Science
Ministry of Education and Science RA, within the frame of Research Project
No. 15 RF-009.

\end{document}